\documentclass[twocolumn,aps,pre,longbibliography,superscriptaddress]{revtex4-2}
\usepackage{amsmath,amssymb,graphicx,dsfont,subfigure}
\usepackage[colorlinks=true, allcolors=blue]{hyperref}

\begin{document}

\title{Anomalous scaling in redirection networks}

\author{Harrison Hartle}
\email{hhartle@santafe.edu}
\affiliation{Santa Fe Institute, Santa Fe, New Mexico 87501, USA}
\author{P. L. Krapivsky}
\email{pkrapivsky@gmail.com}
\affiliation{Department of Physics, Boston University, Boston, Massachusetts 02215, USA}
\affiliation{Santa Fe Institute, Santa Fe, New Mexico 87501, USA}
\author{S. Redner}
\email{redner@santafe.edu}
\affiliation{Santa Fe Institute, Santa Fe, New Mexico 87501, USA}
\author{Yuanzhao Zhang}
\email{yuanzhao.zhang.1@gmail.com}
\affiliation{Santa Fe Institute, Santa Fe, New Mexico 87501, USA}

\begin{abstract}
  In networks that grow by isotropic redirection (IR), a new node selects an initial target node uniformly at random and attaches to a randomly chosen neighbor of the target. The emerging networks exhibit leaf proliferation, in which the number of nonleaves scales sublinearly as $N^\mu$ and the degree distribution has an algebraic tail with exponent $1+\mu$. To understand these mysterious properties, we introduce a class of models with redirection to leaves whenever possible. The resulting networks exhibit qualitatively similar phenomenology to IR networks, but avoid the inherent non-locality of the IR growth rule. These networks admit an analytical description of the leaf degree distribution, from which we extract the exponent $\mu$.
\end{abstract}

\maketitle

\section{Introduction}

In the rich zoo of complex networks, two models stand out for their simplicity and richness: random recursive trees (RRTs)~\cite{Drmota,Flajolet,KRB,Frieze, Hofstad} and preferential attachment (PA) networks~\cite{simon1955class,barabasi1999emergence,KR00,dorogovtsev2000structure,albert2002statistical,Newman-book}. In RRTs, new nodes sequentially join the network by attaching to pre-existing nodes uniformly at random. In PA networks, new nodes attach to pre-existing nodes with a probability that is an increasing function of the degree of the target node. The case where the attachment probability increases linearly with the degree of the target is especially popular because this rule leads to scale-free networks where the degree distribution decays algebraically with a non-universal exponent.

An unexpected feature of PA networks is that it is not necessary to
know the entire set of node degrees to determine where a new node
should attach. The mechanism of redirection~\cite{Kleinberg,KR01,KR14,KR17,Sumpter22,KR24} provides an
algorithmically efficient way to generate preferential attachment by a simple and local deformation of the RRT growth rule. This redirection algorithm gives PA networks that grow by both linear~\cite{KR01} and sublinear~\cite{GR13} preferential attachment. Thus a purely local growth rule implicitly encodes the global information that is ostensibly needed to construct PA networks.

\begin{figure}
    \centering
    \includegraphics[trim = 0 20 0 0, width=0.3\textwidth]{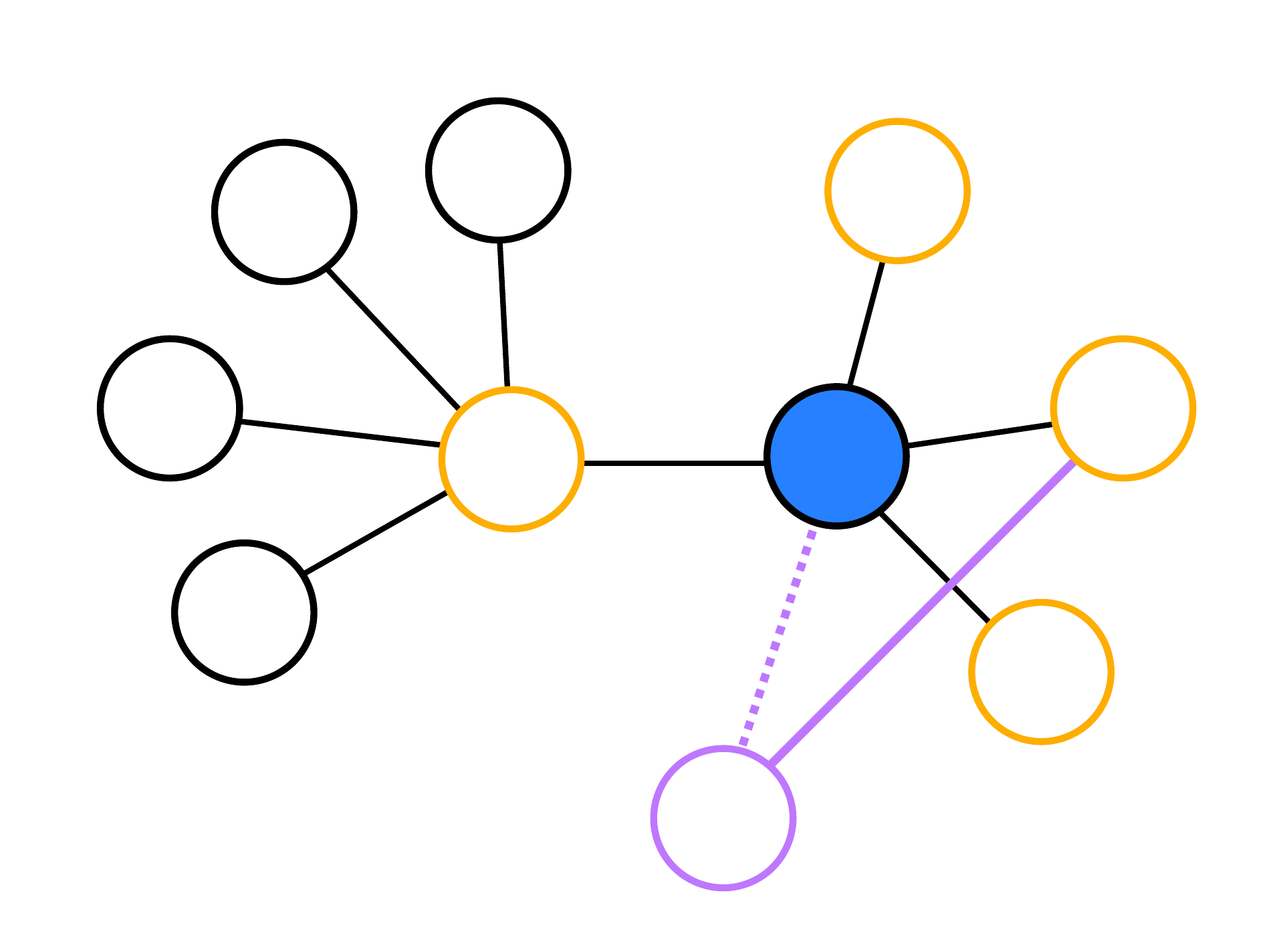}
    \caption{Illustration of isotropic redirection (IR). A new node (purple circle) selects a pre-existing target node (blue) uniformly at random (dashed line). The new node attaches (solid line) to one of the neighbors of the selected node (orange circles) uniformly at random.}
    \label{fig:iso}
\end{figure}

Redirection has even more intriguing consequences for undirected trees grown by the following rule:
\begin{enumerate}
\itemsep -0.5ex
\item Select a pre-existing node uniformly at random.
\item Attach to a random neighbor of the selected node.
\end{enumerate}
This rule is known as isotropic redirection (IR); see Fig.~\ref{fig:iso}. Hereinafter, we assume that each new node attaches only to a single pre-existing node, so that the resulting network is a tree. This isotropic redirection rule leads to highly modular trees in which almost all nodes are leaves (nodes of degree $1$), while the number $\mathcal{C}$ of `core' nodes (nonleaves) scales sublinearly with the total number of nodes $N$: $\mathcal{C}=O(N^\mu)$, with $\mu\approx 0.566$ \cite{KR17}. Moreover, the degree distribution of the core nodes is extraordinarily heavy-tailed:
\begin{align}
  \label{Nk:large}
  N_k\sim \frac{N^\mu}{k^{1+\mu}}\,,
\end{align}
where $N_k$ is the number of nodes of degree $k\geq 2$. The exponent value $\mu\approx 0.566$ seems paradoxical because a powerlaw distribution with exponent $1+\mu<2$ has a diverging first moment, whereas any tree obeys the structural constraint $\sum_{k\geq 1}kN_k=2(N-1)$~\cite{Drmota} implying constant average degree. This apparent contradiction is resolved by the fact that the number of core nodes of any degree scale sublinearly with $N$; that is, $N_k\sim N^\mu$, with exponent shifted by one from the degree exponent. Since the number of leaves $N_1$ is $N_1=N-\mathcal{C}=N-O(N^\mu)$, we have
\begin{equation*}
\sum_{k=1}^N kN_k = N_1+ \sum_{k=2}^N kN_k \sim N+ \sum_{k=2}^N \frac{N^\mu}{k^{\mu}} \sim N\,,
\end{equation*}
so that \eqref{Nk:large} is indeed consistent. This behavior of the degree distribution dramatically contrasts with typical sparse networks in which the degree distribution is extensive, $N_k\sim N/k^\nu$, with $\nu>2$.

In spite of the algorithmic simplicity of the IR growth rule, analytical progress has been stymied by its nonlocal nature. Namely, the probability $w_i$ to attach to a given node $i$ depends on the degrees of all its neighbors:
\begin{equation*}
w_i\propto \sum_{j\sim i}\frac{1}{k_j}\;,
\end{equation*}
where $k_j$ is the degree of node $j$, and $j\sim i$ denotes the neighbors $j$ of node $i$.  Because of this nonlocality, determining the degree distribution requires the joint degree distribution $N_{k,k'}$, which, in turn, requires higher-order degree correlations, etc. What is known rigorously about this model is that the fraction of leaves does approach $1$ as $N\to\infty$~\cite{CC13} and that $\mu \leq 0.9$~\cite{Addario}.

In this work, we avoid the complications caused by the nonlocality of the IR growth rule by introducing closely related growing network models that turn out to have qualitatively similar phenomenology to the IR network. In particular, we consider models in which redirection from nonleaves is only to {\it leaves}, hence avoiding any considerations of degree correlations. These models are analytically tractable via approaches recently developed in the context of growing unlabeled trees~\cite{hartle2025growing}, leaf-based growth models~\cite{hartle26statistics}, and the PA tree~\cite{hartle26}. An essential aspect of these approaches is to consider the \emph{leaf degree} of nodes, i.e., the number of leaves that a given node is attached to \cite{Kaneko01,Lin25}. In particular, we derive and solve recurrences for how the distribution of leaf degree evolves under network growth, akin to conventional considerations of the degree distribution.

In Sec.~\ref{sec:models}, we introduce models of growing trees with leaf-based redirection mechanisms. In Sec.~\ref{sec:mu}, we analyze these models, including analytical determination of the exponent $\mu$ that controls the sublinear core growth and degree distribution tail. In Appendix~\ref{sec:star}, we estimate the asymptotic behavior of the core-size distribution by calculating the probability of star graphs and line graphs. In Appendix~\ref{sec:inf}, we discuss an extreme version of one of our network models that exhibits quite peculiar phenomenology, at the boundary between anomalous and non-anomalous scaling. In Appendix~\ref{sec:2p}, we present a toy two-population model that mimics some of the essential features of the growing network models, providing further insight into the absence of self-averaging in the network models.

\section{Models}
\label{sec:models}

\begin{figure}[ht]
    \centering
    \includegraphics[width=0.3\textwidth]{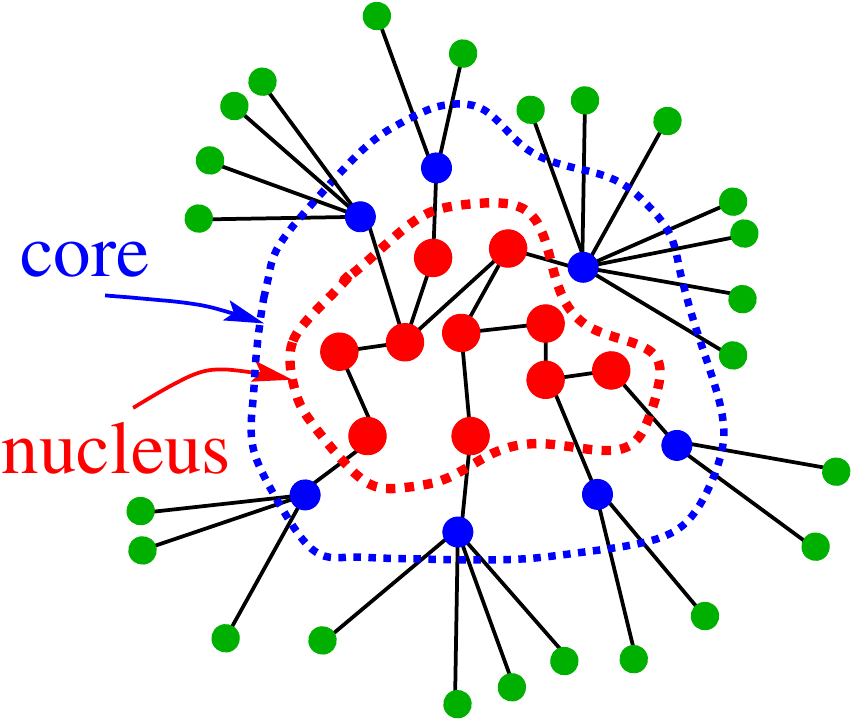}
    \caption{Schematic structure a tree consisting of leaves (green dots), rank-$1$ nodes (blue dots), and rank $>1$ nodes (red dots). The rank $>1$ nodes form the nucleus, and combined with rank-$1$ nodes they form the core.}
    \label{fig:schematic}
  \end{figure}

To define our models, it is useful to partition the network as shown in Fig.~\ref{fig:schematic}. Leaves (green) with degree $1$ form the network periphery. Attached to leaves are rank-$1$ nodes (blue); by definition, rank-$1$ nodes are a distance $1$ from the periphery. We define the nucleus of the network (red outline) as nodes that are more than a unit distance from the periphery; these are nodes with rank $>1$, a.k.a., protected nodes \cite{Prodinger12,Mahmoud15}. The nucleus and rank-$1$ nodes together comprise the network core (blue outline). In terms of this picture, we define the following network growth models.\smallskip

\begin{figure*}[ht]
    \centering
    \subfigure[]{\includegraphics[trim=20 20 20 0,clip,width=0.28\textwidth,angle=90]{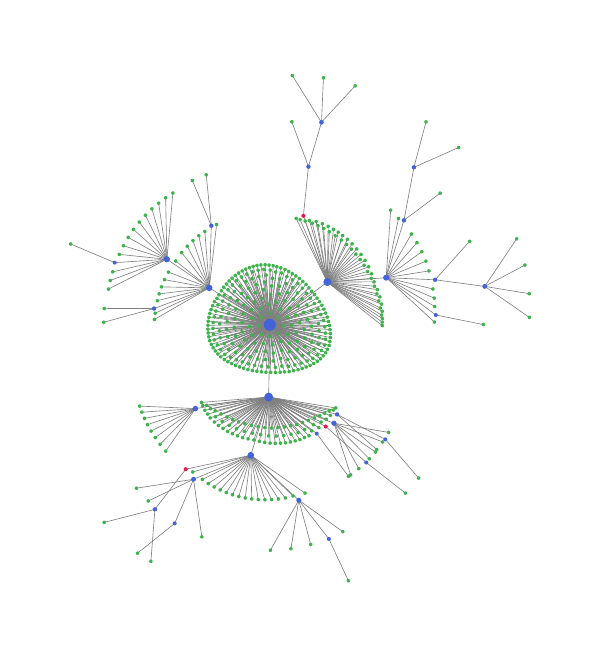}}
    \subfigure[]{\includegraphics[trim=20 20 20 20, clip, width=0.32\textwidth]{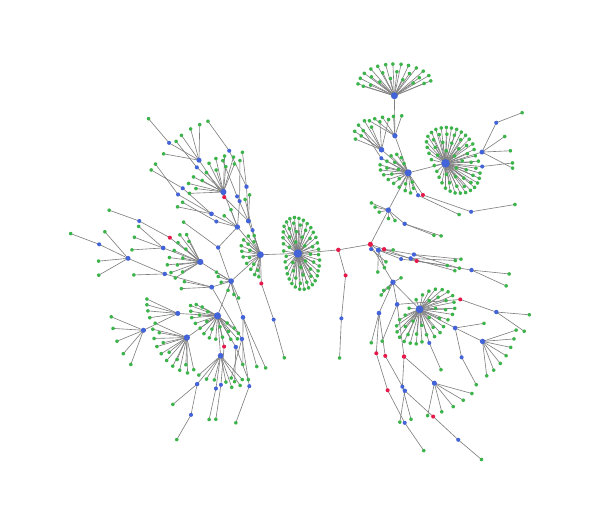}}
    \subfigure[]{\includegraphics[trim=20 20 20 20, clip, width=0.32\textwidth]{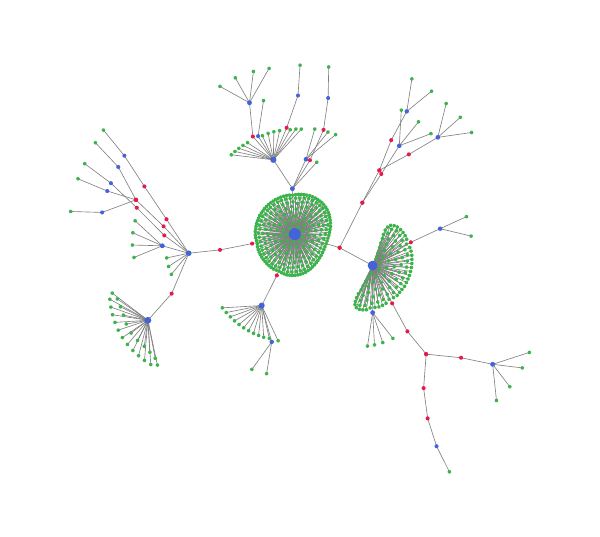}}
    \caption{Random trees grown by: (a) the IR model, (b) the DAN model, and (c) the PAN model. Node colors are as in Fig.~\ref{fig:schematic}.}
    \label{fig:vis}
\end{figure*}

\smallskip\noindent\textbf{Direct attachment to the nucleus (DAN):}
\vspace{-2pt}
\begin{enumerate}
  \itemsep -0.35ex
\item If a leaf is selected, attach to its rank-$1$ neighbor.
\item If a rank-$1$ node is selected, attach to its leaf.
\item {\it If a nucleus node is selected, attach to this node.}
\end{enumerate}

\smallskip\noindent\textbf{Prohibited attachment to the nucleus (PAN):}
\begin{enumerate}
  \itemsep -0.35ex
\item If a leaf is selected, attach to its rank-$1$ neighbor.
\item If a rank-$1$ node is selected, attach to its leaf.
\item {\it If a nucleus node is selected, reject this event.}
\end{enumerate}

Both processes resemble the IR growth rule, with two modifications: In rule $2$, redirection from rank-$1$ nodes is always to leaves in both the DAN and PAN models, whereas in IR, redirection is isotropic among all neighboring leaves and core nodes; in rule 3, the attachment does not involve redirection. In both the DAN and PAN models, if nucleus node is initially selected, redirection to a neighboring node is not allowed, instead undergoing direct attachment (DAN) or rejecting the event (PAN). In the IR model, redirection from a core node can be to another core node; this is the process that leads to higher-body correlations in the underlying rate equations for the IR model. Because core-to-core redirction is forbidden in both the DAN and PAN models, they turn out to be analytically tractable while still exhibiting qualitatively similar behavior to the IR model.

The DAN and PAN models can be subsumed into a general class of models that we term \emph{variable attachment to the nucleus}, VAN$(\omega)$. Here, nucleus nodes are assigned a weight $\omega$, and all other nodes have a weight of $1$. For leaves and rank-$1$ nodes, the redirection mechanism is exactly the same as in the DAN and PAN models. If a nucleus node is selected as the target, which happens with probability $\omega/(\omega \mathcal{N} + N-\mathcal{N})$, where $\mathcal{N}$ is the number of nodes in the nucleus, the new node attaches to this target. The DAN model corresponds to $\omega=1$, and the PAN model is recovered when $\omega=0$. In the extreme case of $\omega=\infty$, each nucleus node that arises is immediately attached to in the subsequent growth step, becoming rank-$1$.

Networks that are generated by the IR, DAN, and PAN mechanisms are indeed similar in appearance and in structural details (Fig.~\ref{fig:vis}), and exhibit the following essential features:
\begin{itemize}

\item\textbf{Leaf proliferation:} Almost all nodes are leaves, while the
  remaining core of the network grows as
  \begin{equation}
  \label{eq:core_scaling}
      \mathcal{C}=N-N_1\sim N^\mu,
    \end{equation}
    with $\mu<1$.
\item\textbf{Non-self-averaging of $\mathcal{C}$:} The size of the core is
  broadly distributed.  For $N, \mathcal{C}\to\infty$, with
  $z=\mathcal{C}/\langle \mathcal{C}\rangle$ finite, the probability
  distribution of core size, $P_N(C)$, approaches the scaling form
\begin{equation}
\label{Pz:scaled}
P_N(C)=\langle\mathcal{C}\rangle^{-1}\mathcal{P}(z),
\qquad z=C/\langle\mathcal{C}\rangle\,.
\end{equation}
Similar non-self-averaging holds for the variables $N_k$ ($k>1$), $M_\ell$ ($\ell>0$), and $\mathcal{N}$.

\item\textbf{Algebraic decay of $N_k$ and $M_\ell$:} The average values of $N_k$ and $M_\ell$ satisfy
\begin{align}
\begin{split}
  \label{ckml}
\langle N_k\rangle& \sim N^\mu k^{-1-\mu},        \qquad k\gg 1, \\
\langle M_\ell\rangle& \sim N^\mu \ell^{-1-\mu}, \qquad \ell\gg 1.
\end{split}
\end{align}
for the same exponent $\mu<1$ as in \eqref{eq:core_scaling}. The equality of the degree and leaf degree tail exponents is a consequence of the leaf proliferation phenomenon.

\end{itemize}

Our main results determine the exponent $\mu$: for the DAN model, we show that $\mu$ is the root of the transcendental equation
\begin{subequations}
\label{mu-both}
\begin{equation}
\label{zeta:ULR}
1-\mu^2 = \int_0^1 dx\,x^\mu\, e^{x-1}\;,
\end{equation}
while for the PAN model, we show that $\mu$ is instead determined by
\begin{equation}
\label{zeta:PLR}
1-\mu = \int_0^1 dx\,x^\mu\, e^{x-1}.
\end{equation}
\end{subequations}
The numerically exact values for the exponent $\mu$ are obtainable by solution of \eqref{mu-both}. To six decimal places, the values are $\mu\approx 0.771\, 192$ (DAN) and $\mu\approx 0.549\, 735$ (PAN). Additionally, we determine the tail behavior of the scaled distribution of core size \eqref{Pz:scaled}, namely,
\begin{subequations}
\label{eq:Pz_asymp}
\begin{equation}
\label{Pz:small}
\mathcal{P}(z) \sim z^{-1+1/\mu}\qquad z\to 0\,,
\end{equation}
while for $z\to\infty$
\begin{equation}
\label{Pz:large}
-\log \mathcal{P}(z) \sim 
\begin{cases}
z^{1/(1-\mu)} \log z  & (\text{DAN})\\
z^{1/(1-\mu)}           & (\text{PAN})\,,
\end{cases}
\end{equation}
\end{subequations}
as derived in Appendix~\ref{sec:star}. Furthermore, this work clarifies the nature of self-averaging in these models: {\it ratios} of the form $\mathcal{N}/\mathcal{C}$, $M_\ell/\mathcal{C}$ ($\ell>0$), and $N_{k}/\mathcal{C}$ ($k>1$) are asymptotically self-averaging---approaching nonrandom limiting values, denoted $q_0$, $q_\ell$, and $c_k$, respectively---despite the absolute quantities in the numerators and denominators being non-self-averaging.

\section{Theoretical analysis}
\label{sec:mu}

In this section, we establish recurrences for the leaf degree distributions of the models introduced in Sec.~\ref{sec:models}, resulting in the implicit formulae \eqref{mu-both} for the anomalous exponent $\mu$ that quantifies the core size growth and the degree distribution tail. The variables of interest are $\mathcal{N}$ (the number of nodes in the nucleus), $N_1=N-\mathcal{C}$ (the number of leaves), and $M_\ell$ (the number of nodes with leaf degree $\ell$, for $\ell=1,2,3,\ldots$). Nodes with leaf degree zero consist of nodes in the nucleus and the leaves themselves, so that $M_0 = N_1 + \mathcal{N}$.  Thus,
the quantities $(M_0,N_1,\mathcal{N})$ are not all independent and we can
choose any pair of them to characterize the network. We choose
$(N_1,\mathcal{N})$, together with $M_\ell$ for $\ell>0$, as the fundamental variables in what follows. It will also be useful to exploit the sum rules obeyed by the leaf degree distribution:
\begin{align}
  \sum_{\ell\geq 0} M_\ell=N, \qquad  \sum_{\ell\geq 1} \ell M_\ell=N_1\,,
\end{align}
The first equation states that the number of nodes with any leaf degree must equal the total number of nodes; the second equation expresses the total number of leaves through the leaf degree distribution.

\begin{figure}[ht]
\includegraphics[width=0.44\textwidth]{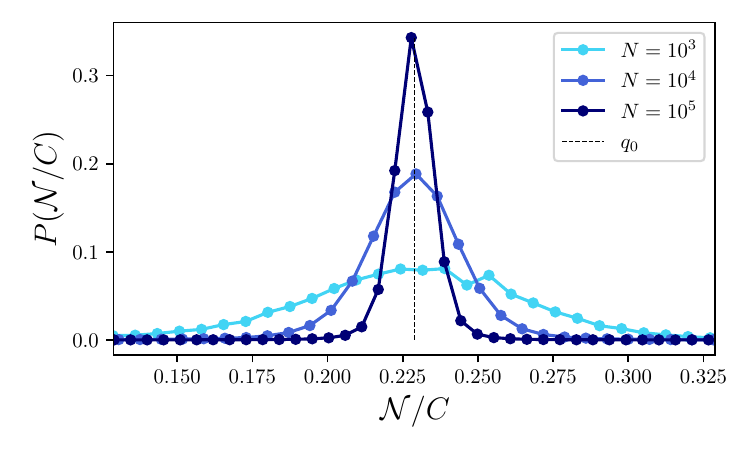}
\includegraphics[width=0.44\textwidth]{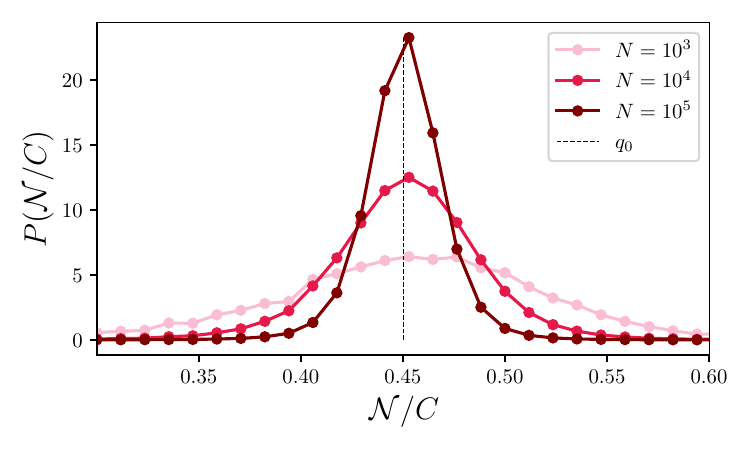}
\caption{Illustration of self-averaging for the nucleus to core ratio. The scaled probability distribution $P(\mathcal{N}/\mathcal{C})$ for the ratio of nucleus to core size \emph{for a given network realization}, $\mathcal{N}/\mathcal{C}$, as a function of the number of nodes $N$. The distribution converges to a delta function that is located at the deterministic value $q_0$. Shown are data for the DAN model (upper panel) and the PAN model (lower panel).}
\label{fig:selfavg}
\end{figure}

The variables $M_\ell$, $N_k$, $\mathcal{N}$, and $\mathcal{C}$ remain asymptotically {\it non}-self-averaging, namely, relative fluctuations between different network realizations do not vanish in the $N\to\infty$ limit. On the other hand, we find that {\it ratios} of the above basic random quantities are asymptotically self-averaging. We analyzed some representative such quantities, particularly $\mathcal{N}/\mathcal{C}$, $M_\ell/\mathcal{C}$ with small $\ell>0$, and $N_k/\mathcal{C}$ with small $k>1$. Fig.~\ref{fig:selfavg} illustrates the asymptotic self-averaging of the ratio of nucleus size $\mathcal{N}$ to core size $\mathcal{C}$ for both the DAN and PAN models. The high correlation of $M_{\ell}$ with $\mathcal{C}$ results in ratios which are asymptotically deterministic:
\begin{align}
\label{SA}
\frac{\mathcal{N}}{\mathcal{C}} \rightarrow  q_0, \qquad \frac{M_\ell}{\mathcal{C}} \rightarrow  q_\ell.
\end{align}
This notion of asymptotic self-averaging is the basis of the following analysis. We also observe convergence of the degree distribution within the core of the form $N_k/\mathcal{C}\rightarrow c_k$ for $k>1$, but our analysis pertains to the leaf degree distribution.

In the DAN and PAN models, and more generally the
VAN($\omega$) model class, we will see that the limiting ratios $\{q_\ell\}_{\ell\ge 0}$ satisfy closed and solvable recurrences.  In contrast, for the IR model, the fractions $q_\ell$ do not satisfy a closed recurrence. Moreover, the fractions $c_k=N_k/\mathcal{C}$ do not satisfy closed recurrences for any of these models. This simplification in the DAN and PAN models highlights the importance of focusing on the leaf degree as a fundamental dynamical variable. We note that closed two-index recurrences for the joint distribution of degree and leaf degree \cite{hartle26} (within the core) are obtainable for the DAN, PAN, and VAN($\omega$) model, but appear not to be for IR.

\subsection{DAN model}
\label{ssec:dan}

For the DAN model, the quantities $(N_1,\mathcal{N})$ evolve according to
\begin{equation}
\label{N1P}
\frac{dN_1}{dN} = \frac{N_1+\mathcal{N}}{N},
\qquad \frac{d\mathcal{N}}{dN} = \frac{M_1-\mathcal{N}}{N}\,.
\end{equation}
In the first equation, the number of leaves increases by $1$ whenever a leaf or nucleus node is initially selected.  For the former choice, the new node attaches to a rank-$1$ node while for the latter, direct attachment to the nucleus node occurs.  In both cases, the number of leaves increases by $1$. For the second equation, whenever a new node directly attaches to a nucleus
node, the latter becomes a rank-$1$ node and is removed from the nucleus. When a new node first selects a rank-$1$ node with a
single leaf, redirection to this leaf occurs.  In this case, the number of nucleus nodes increases by $1$.

\begin{figure}[ht]
\subfigure[]{\includegraphics[width=0.125\textwidth]{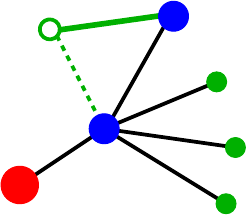}}\qquad\qquad
\subfigure[]{\includegraphics[width=0.125\textwidth]{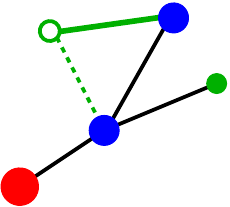}}\\
\subfigure[]{\includegraphics[width=0.15\textwidth]{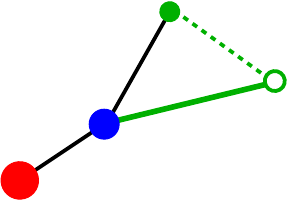}}\qquad\qquad
\subfigure[]{\includegraphics[width=0.125\textwidth]{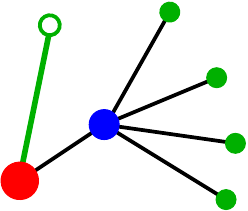}}
\caption{The four processes that contribute to the change in $M_1$. The new node, necessarily a leaf, is indicated by the open green circle. The dashed green line indicates the initially selected node and the solid green line indicates the attachment event. The blue node is rank-$1$ and the red node is protected (other neighbors not displayed). In (a) and (b) the new attachment turns a leaf into a rank-$1$ node.}
\label{fig:M1schematic}
\end{figure}

The number of nodes with leaf degree equal to 1, $M_1$, grows with $N$ according to
\begin{align}
\label{M1}
\frac{dM_1}{dN} =  \frac{N-N_1-2M_1+M_2}{N} \;.
\end{align}
There are four distinct processes that contribute to $dM_1/dN$, as
illustrated in Fig.~\ref{fig:M1schematic}. (a) A rank-$1$ node with leaf degree $\geq 3$ is selected and redirection to a leaf occurs; here $M_1$ increases by $1$. (b) A rank-$1$ node with leaf degree $2$ is selected and redirection to leaf occurs; here $M_1$ increases by $2$. (c) A leaf that is attached to a rank-$1$ node is selected and redirection to the rank-1 node occurs; here $M_1$ decreases by $1$. (d) A nucleus node is selected and attachment to this node occurs; $M_1$ increases by $1$. Assembling these
processes, using the sum rule $\sum_\ell M_\ell=N$, as well as
$M_0=\mathcal{N}+N_1$, then gives the right-hand side of Eq.~\eqref{M1}.

Finally, the leaf degree distribution $M_\ell$ with $\ell\geq 2$ satisfy the infinite system of equations
\begin{align}
\label{M-ell}
\frac{dM_\ell}{dN}  =  \frac{(\ell-1)M_{\ell-1}+M_{\ell+1}-(\ell+1)M_\ell}{N}\; .
\end{align}
The three processes that contribute to $dM_\ell/dN$ are the following: (a) Select one of the $\ell-1$ leaves neighboring a rank-$1$ node with leaf degree $\ell-1$ and attach to that rank-$1$ node; in this case, $M_\ell\to M_\ell+1$. (b) Select a rank-$1$ node of leaf degree $\ell+1$ and attach to one of the leaves of this node; again, $M_\ell\to M_\ell+1$. (c) Either select one of the $\ell$ leaves whose rank-$1$ neighbor has leaf degree $\ell$ and attach to the rank-$1$ node, or select a rank-$1$ node of leaf degree $\ell$ and attach to one of its leaves; the result is $M_\ell\to M_\ell-1$.

\begin{figure}[ht]
    \centering
    \includegraphics[width=0.93\linewidth,trim=20 20 0 0]{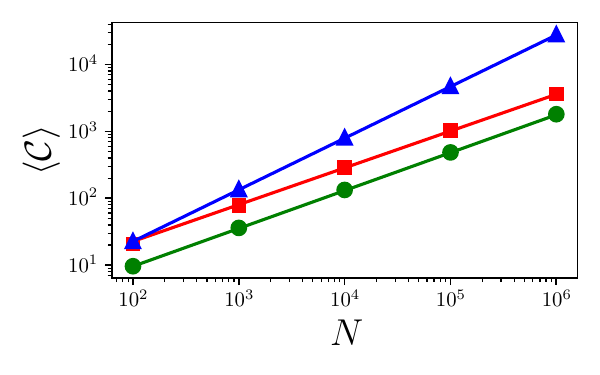}
    \caption{The average number of core nodes $\langle \mathcal{C}\rangle$ as a function of the number of network nodes from $N=10^2$ to $N=10^6$ for the DAN
      ($\blacktriangle$) and PAN ($\blacksquare$) models, as well as for the IR
      model ($\bullet$). Each point represents the mean of $\mathcal{C}$ over 2400 independent realizations. A nonlinear least-squares fit provides accurate estimates for the exponents: $\mu\approx 0.569$ for the IR model, and $\mu\approx 0.773$ and $0.545$ for DAN and PAN, respectively, aligning approximately with the analytically computed values.}
    \label{fig:n1_growth}
\end{figure}

\begin{figure}[ht]
\includegraphics[trim=10 30 0 0,clip, width=0.99\linewidth]{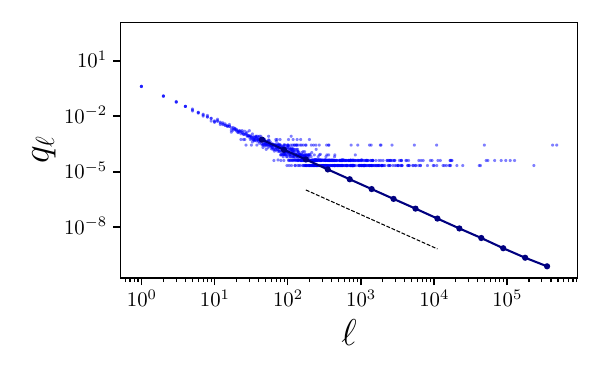}
\includegraphics[trim = 10 0 0 0, width=0.99\linewidth]{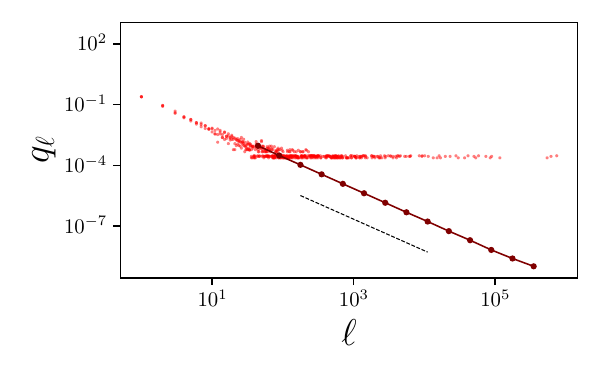}
\caption{Leaf degree distributions of the DAN model (upper panel) and PAN model (lower panel). The colored dots are the raw data (values of $M_\ell/\mathcal{C}$) for three graphs of size $N=10^6$; the dotted curve represents log-binned data from $4800$ graphs of size $N=10^6$; the black dashed lines indicate pure powerlaw decay with slopes predicted by our theory: $1+\mu_{\mathrm{DAN}}\approx 1.771$ and $1+\mu_{\mathrm{PAN}}\approx 1.549$, respectively.}
\label{fig:leaftail}
\end{figure}

A linear ansatz of the form $M_\ell\simeq m_\ell N$ results in the trivial condition $m_\ell=0$ for all $\ell>0$, and $m_0=1$ as indicative of leaf proliferation. Instead, a sublinear growth ansatz is taken; Fig.~\ref{fig:n1_growth} confirms the sublinear growth of the average size of the core, $\langle \mathcal{C}\rangle\sim N^\mu$, with numerically estimated exponent $\mu$ close to the values we analytically derive in what follows. Introducing the sublinear scaling ansatz and limiting fractions $q_\ell$ of \eqref{SA} into \eqref{M-ell}, the system reduces to recurrence
\begin{align}
\label{q-ell}
\mu q_\ell & =  (\ell-1)q_{\ell-1}+q_{\ell+1}-(\ell+1)q_\ell, \quad \ell\geq 2\,,
\end{align}
while Eqs.~\eqref{N1P} and \eqref{M1} become 
\begin{align}
  \label{q01}
\begin{split}
& q_0 = 1-\mu, \quad q_1=1-\mu^2,\\
& \mu q_1  =  1-2q_1+q_2.
\end{split}
\end{align}
Substituting \eqref{q01} into \eqref{q-ell}, one can express $q_\ell$ as
polynomials in $\mu$ of degree $\ell+1$.  For example, the first few of these
polynomials are
\begin{equation*}
\begin{split}
q_2 &= 1+\mu-2\mu^2-\mu^3,\\
q_3 &= 2+4\mu-4\mu^2-5\mu^3 - \mu^4,\\
q_4 &= 6+16\mu-8\mu^2-22\mu^3 -9\mu^4 - \mu^5,
\end{split}
\end{equation*}
etc. In general, solutions to \eqref{q-ell} grow factorially with $\ell$. Indeed, Eq.~\eqref{q-ell} becomes $q_{\ell+1}=(\ell+\mu+1)q_\ell$ to leading order. This fact suggests that we seek a solution of the form
$q_\ell= Q_\ell\Gamma(\ell+\mu+1)$. Substituting this ansatz into
Eq.~\eqref{q-ell}, we find a simple recurrence for $Q_\ell$ that has an asymptotic solution $Q_\ell\simeq C(\mu)/(\ell+\mu)$.  Hence the growth of $q_\ell$ is indeed factorial: $q_\ell\simeq C(\mu)\Gamma(\ell+\mu)$. For specific values of $\mu$, the amplitude $C(\mu)$ vanishes, and a second linearly-independent solution of the linear three-term recurrence \eqref{q-ell} determines the asymptotic. The asymptotic of this {\em minimal at infinity} solution turns out to be algebraic, so it can be extracted using a continuum framework where \eqref{q-ell} simplifies to $\frac{d}{d\ell}(\ell q_\ell)=-\mu q_\ell $ leading to $q_\ell\sim \ell^{-1-\mu}$ for $\ell\gg 1$. Fig.~\ref{fig:leaftail} shows the tail behavior of $q_\ell$, corroborating the predicted algebraic decay.

Fortunately, the explicit solutions to the generic three-term recurrences of the form
$q_{\ell+1}+a_\ell q_\ell+b_\ell q_{\ell-1}=0$ that are minimal at infinity
are known \cite{Rec3}.  These solutions exhibit a remarkable representation
in terms of the continued fractions \cite{Rec3}. In particular, for the
generic three-term recurrence, the ratio $q_2/q_1$ is
\begin{equation}
\label{CF:ab}
\frac{q_2}{q_1} = - \cfrac{b_2}{a_2-\cfrac{b_3}{a_3-\cfrac{b_4}{a_4-\cdots}}}~.
\end{equation}
Recalling that $q_1=1-\mu^2$ and $q_2= 1+\mu-2\mu^2-\mu^3$ in the DAN model,
we find that the ratio $q_2/q_1$ is
\begin{equation*}
\frac{q_2}{q_1} =  2+\mu-\frac{1}{1-\mu^2}\;.
\end{equation*}
Equating this result to \eqref{CF:ab} which we specialize to $a_\ell = -(\ell+1+\mu)$ and $b_\ell = \ell-1$ as per the three-term recurrence \eqref{q-ell}, we obtain
\begin{equation}
\label{zeta:ULR-F}
1-\mu^2 = F(\mu)\,,
\end{equation}
where we introduce the shorthand notation
\begin{equation}
\label{F:def}
\frac{1}{F(\mu)}= 2+\mu-\cfrac{1}{3+\mu-\cfrac{2}{4+\mu-\cfrac{3}{5+\mu-\cdots}}}~.
\end{equation}

With the help of the handbook \cite{Cont_Fractions}, we can express the continued fraction in terms of an integral representation. We set $a=1+\mu$ and $z=-1$ in identity (12.6.30) of \cite{Cont_Fractions}, and then use the integral representation of the incomplete gamma function $\gamma(a,z)=\gamma(1+\mu,-1)$ in the identity (12.6.30) to give
\begin{equation*}
F(\mu) = \int_0^1 dx\,x^\mu e^{x-1}\;.
\end{equation*}
Combining this with \eqref{zeta:ULR-F}, we obtain the
transcendental equation for $\mu$
\begin{equation}
  \label{mu-final}
1-\mu^2 = \int_0^1 dx\,x^\mu\, e^{x-1}\;.
\end{equation}
The numerically exact value of the exponent $\mu$ to six decimal places is $\mu\approx 0.771\,192$.

\subsection{PAN model}

For the PAN model, the quantities $M_\ell$ with $\ell\geq 2$ satisfy equations similar to Eqs.~\eqref{M-ell}, except that the denominator $N$ should now be replaced by $N-\mathcal{N}$. This difference can be ignored in the $N\to\infty$ limit, so that Eqs.~\eqref{q-ell} remain valid. The primary differences with the DAN model are the equations \eqref{N1P} and \eqref{M1} for $N_1$, $\mathcal{N}$, and $M_1$. First, we note that in each node addition, the target node is selected uniformly at random among all $N-\mathcal{N}$ non-nucleus nodes. By enumerating all the ways that $N_1$ and $\mathcal{N}$ can change when a new node joins the network, we arrive at the rate equations
\begin{equation}
\label{N1P:P}
\frac{dN_1}{dN} = \frac{N_1}{N-\mathcal{N}},
\qquad \frac{d\mathcal{N}}{dN} = \frac{M_1}{N-\mathcal{N}}\;.
\end{equation}
In the equation for $dN_1/dN$, the number of leaves increases by 1 whenever the new node first selects a leaf and then attaches to its rank-$1$ neighbor. In the equation for $d\mathcal{N}/dN$, the size of the nucleus increases by $1$ whenever the new node selects a rank-$1$ node of leaf degree $1$ and then attaches to its leaf.

We determine the rate equation for $M_1$ by the same detailed bookkeeping of the processes that contributed to the change in $M_1$ in the DAN model (Sec.~\ref{ssec:dan}). We thereby obtain
\begin{align}
\label{M1:P}
\frac{dM_1}{dN} =  \frac{N-\mathcal{N}-N_1-2M_1+M_2}{N-\mathcal{N}} \;.
\end{align}
This is nearly the same as Eq.~\eqref{M1} for $M_1$ in the DAN model except there is now an extra loss term $-\mathcal{N}$. This term reflects the fact that attachment to a nucleus node is not allowed (the event in Fig.~\ref{fig:M1schematic}(d) is forbidden). The corresponding equation for $M_\ell$ with $\ell\geq 2$ is
\begin{equation}
\label{eq:dotMl}
\frac{d M_{\ell}}{dN}=\frac{(\ell-1)M_{\ell-1}+M_{\ell+1}-(\ell+1)M_\ell}{N-\mathcal{N}}.
\end{equation}

One may again demonstrate directly that a linear scaling ansatz yields the trivial outcome $M_\ell/N\rightarrow 0$ for $\ell>1$ indicating sublinear growth, alongside $N_1/N\rightarrow 1$ indicating leaf proliferation; we proceed again with the ansatz $N_1=N-O(N^\mu)$ and $M_\ell\sim N^\mu q_\ell$. From Eqs.~\eqref{N1P:P} and \eqref{M1:P}, the analogs of Eqs.~\eqref{q01} are
\begin{align}
\begin{split}
\label{q01:P}
   q_0 &= 1-\mu\, ,\\
  q_1&=\mu-\mu^2\, ,\\
\mu q_1  &=  1-q_0-2q_1+q_2\,.
\end{split}
\end{align}
Therefore $q_2=\mu-\mu^2-\mu^3$, and now using \eqref{q01:P},
the ratio $q_2/q_1$ is
\begin{align*}
  \frac{q_2}{q_1} = 2+\mu-\frac{1}{1-\mu}\,.
\end{align*}
Following the same steps as those used for the DAN model, we obtain the transcendental equation for $\mu$,
\begin{equation}
\label{zeta:PLR-F}
1-\mu = F(\mu) =  \int_0^1 dx\,x^\mu e^{x-1}\,.
\end{equation}
The numerically exact value of the exponent $\mu$, up to six decimal places, is $\mu\approx 0.549\,735$.

\subsection{VAN$(\omega)$ model}
\label{sec:omega}

\begin{figure}
    \centering
    \includegraphics[width=0.85\linewidth]{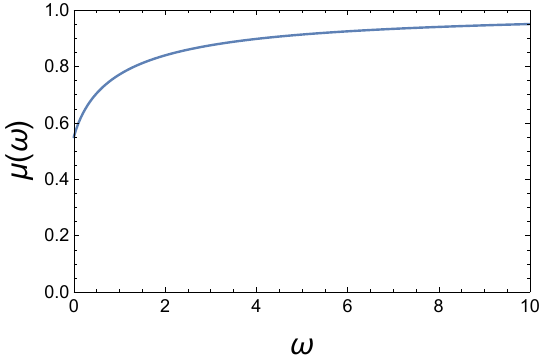}
    \caption{The scaling exponent $\mu(\omega)$ of the VAN($\omega$) model. It monotonically increases from $\mu(0)=\mu_{\mathrm{PAN}}\approx 0.549$ to $\mu(\infty)=1$ with asymptotic behavior \eqref{mu:omega-inf}.}
    \label{fig:mu_of_omega}
\end{figure}

For the VAN$(\omega)$ model, the growth of the number of leaves, the nucleus, and the number of nodes of each leaf degree value $\ell>0$ are accounted for by the rate equations
\begin{align}
\begin{split}
  \label{N1:omega}
&\frac{dN_1}{dN} = \frac{N_1+\omega \mathcal{N}}{N+(\omega-1)\mathcal{N}}\, , \\[1mm]
&\frac{d\mathcal{N}}{dN} = \frac{M_1- \omega \mathcal{N}}{N+(\omega-1)\mathcal{N}}\,,  \\[1mm]
&\frac{dM_1}{dN} =  \frac{N+(\omega-1)\mathcal{N}-N_1-2M_1+M_2}{N+(\omega-1)\mathcal{N}}\, , \\[1mm]
&\frac{dM_\ell}{dN} =  \frac{(\ell-1)M_{\ell-1}+M_{\ell+1}-(\ell+1)M_\ell}{N+(\omega-1)\mathcal{N}}\, .
\end{split}
\end{align}
The same approach as in the DAN and PAN models yields the same universal result $q_0 = 1-\mu$ for the fraction of nucleus nodes in the core, while the ratios $q_1/q_0$ and $q_2/q_1$ are given by
\begin{equation}
\label{q12:omega}
\frac{q_1}{q_0}=\mu+\omega, \quad
\frac{q_2}{q_1}=2+\mu-\frac{\omega+(1-\omega)\mu}{(1-\mu)(\mu+\omega)}\;.
\end{equation}
Finally, the exponent $\mu$ is determined from the root of the transcendental
equation
\begin{equation}
\label{mu:omega}
\frac{(1-\mu)(\mu+\omega)}{\omega+(1-\omega)\mu} = \int_0^1 dx\,x^\mu e^{x-1}\;.
\end{equation}
The exponent $\mu=\mu(\omega)$ is a monotonic function of $\omega$ that increases from $\mu(0)=\mu_\text{PAN}\approx 0.549$ to $\mu(\infty)=1$. The asymptotic behavior of the exponent in the $\omega\to \infty$ limit
\begin{equation}
\label{mu:omega-inf}
1-\mu(\omega)\simeq \frac{1}{(e-1)\omega}\,
\end{equation}
follows from \eqref{mu:omega}; Fig.~\ref{fig:mu_of_omega} displays $\mu(\omega)$.

\subsection{VAN($\infty$) model}

The limiting model VAN($\infty$) exhibits interesting marginal behavior. In this model, any protected node arising is immediately attached to in the following growth step. The exponent $\mu(\infty)=1$ leads to a quadratically decaying leaf degree tail, $q_\ell\sim \ell^{-2}$. The core grows only logarithmically slower than $N$:
\begin{equation}
\mathcal{C} \sim \frac{N}{\log N},
\end{equation}
illustrating that VAN($\infty$) resides at the boundary between anomalous and non-anomalous scaling. 

The governing recurrence for $\ell\ge 2$ arises by taking $\mu=1$ in \eqref{q-ell},
\begin{align}
0 =  (\ell-1)q_{\ell-1}-(\ell+2)q_\ell +q_{\ell+1}\,.
\end{align}
We derive an exact integral expression for $q_\ell$, namely,
\begin{equation}
    q_\ell=\frac{1}{e-1}\int_0^1dt \, e^{1-t}(1-t)t^{\ell-1}.
\end{equation}
Details of the VAN($\infty$) are provided in Appendix~\ref{sec:inf}.

\section{Discussion}
\label{sec:sum}

We introduced a class of network growth models based on redirection in which redirection from nonleaves to other nonleaves is forbidden. The resulting networks exhibit leaf proliferation and anomalously broad degree and leaf degree distributions---bearing qualitative similarity to isotropic redirection (IR)---yet, these models proved more tractable analytically than IR. The IR growth rule is exceedingly simple: pick a node uniformly at random and attach to a random neighbor of this selected node. Despite this simplicity, the IR network has resisted analytical understanding because the growth rule turns out to be nonlocal in character. The models we introduced retain the spirit and phenomenology of the IR network, but the growth rule remains local in character so that the resulting network is analytically tractable. In particular, this locality arises due to redirection from nonleaf nodes always being to leaves; for nucleus nodes, no such redirection is possible, so we considered the case of direct attachment to the nucleus (DAN) and prohibited attachment to the nucleus (PAN). We also introduced a class of models VAN($\omega$) by assigning a weight $\omega$ to nucleus attachment, which includes the DAN and PAN models as special cases, and has interesting limiting case VAN($\infty$).

Our main results are (a) analytical predictions of the exponent $\mu$ that controls the growth rate of the core: $\mathcal{C}\sim N^\mu$, (b) the ultrabroad leaf degree distribution of nodes in the core: $M_\ell\sim N^\mu \ell^{-(1+\mu)}$, (c) the tail behavior of the rescaled core-size distribution $\mathcal{P}(z)$ where $z=\mathcal{C}/\langle\mathcal{C}\rangle$, and (d) that the ratios $M_{\ell}/\mathcal{C}$ are self-averaging despite the numerator and denominator being individually non-self-averaging. These behaviors contrast dramatically with standard scale-free network models in which degree distributions have a powerlaw exponents strictly greater than $2$. More generally, this work has provided a modeling setting and tractable analytical framework for exploring anomalous scaling and leaf proliferation in growing random trees.

We have not explicitly computed $c_k$, the degree distribution within the core, which does not satisfy a closed recurrence akin to that for $q_\ell$. However, by leaf proliferation, the degree distribution is also powerlaw with an identical exponent, i.e., $c_{k}\sim k^{-1-\mu}$. Despite $c_{k}$ not satisfying a closed recurrence in VAN($\omega$), the joint distribution of degree and leaf degree within the core $c_{k,\ell}$ does, providing a path to compute $c_k$. Similarly, the joint distribution of degree and leaf degree was used to obtain the leaf degree distribution in the PA tree \cite{hartle26}, where it does not satisfy a closed recurrence. Numerous additional questions are of interest these models---for instance, a more detailed characterization of the nucleus, such as the number of components it has, or its distribution of degrees, $c_{k,0}$. In addition, it would of interest to obtain concentration and normality results, akin to those previously established for RRTs and PA \cite{Bollobas01,Janson05}, but in the context of the unusual form of self-averaging arising in these models.

\bigskip\noindent

\appendix

\section{Star and line graphs and the scaled core-size distribution $\mathcal{P}(z)$}
\label{sec:star}

\begin{figure}
    \centering
    \includegraphics[width=0.9\linewidth,trim=0 30 0 0,clip]{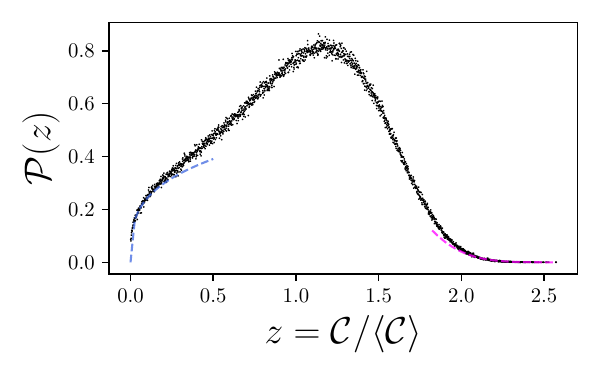}
\includegraphics[width=0.9\linewidth]{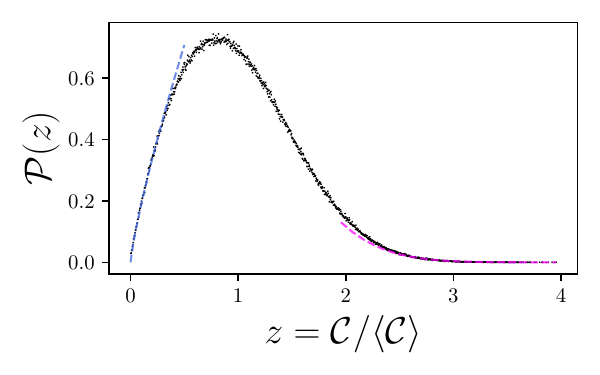}
    \caption{Probability density of rescaled core size $\mathcal{P}(z)$ with $z=C/\langle C\rangle$, for graphs of size $N=10^4$ from the DAN model (upper panel) and PAN model (lower panel), with data displayed from $2.4\times 10^6$ independent realizations. Blue and pink dashed curves represent the asymptotics \eqref{eq:Pz_asymp} with roughly estimated parameters.}
    \label{fig:Pz}
\end{figure}

 We confirmed that the probability distribution of the core size approaches the scaling form
\begin{equation}
\label{eq:P_N_C_scaled}
    P_N(C)=\langle \mathcal{C}\rangle^{-1}\mathcal{P}(C/\langle \mathcal{C}\rangle).
\end{equation}
The scaled distribution $\mathcal{P}(C/\langle \mathcal{C}\rangle)$ is asymptotically stationary and broad, namely, it does not approach a delta function that would indicate asymptotic self-averaging; see Fig.~\ref{fig:Pz}. To obtain the small-$z$ behavior \eqref{Pz:small} and large-$z$ behavior \eqref{Pz:large} of $\mathcal{P}(z)$, we analyze the extremal probabilities of star graphs and line graphs.

A star graph contains a single hub while all other nodes are leaves.  For the RRT, the natural initial condition is $N=3$, which is necessarily a star. Thus $S_3=1$. The probability $S_{N+1}$ to create a star of $N+1$ nodes from a star of $N$ nodes satisfies $S_{N+1}=S_N/N$, from which
\begin{equation}
\label{star:RRT}
S_N = \frac{2}{(N-1)!}\, .
\end{equation}
Similarly for PA networks, the probability to create a star of $N+1$ nodes from a star of $N$ nodes satisfies
$S_{N+1}=S_N/2$, which yields
\begin{align}
\label{star:PA}
  S_N = \frac{1}{2^{N-3}}\, ..
\end{align}
On the other hand, for the DAN, PAN, and VAN($\omega$) models with arbitrary $\omega\geq 0$, as well as for the IR network, the probability to generate a star graph satisfies the recurrence $S_{N+1} = \frac{N-1}{N}S_N$, from which
\begin{align}
\label{star:DAN}
  S_N = \frac{2}{N-1}\,. 
\end{align}
We now use this simple result to determine the small-$z$ tail of the core-size distribution. Specializing \eqref{Pz:scaled} to $\mathcal{C}=1$ gives $P_N(1)=\langle\mathcal{C}\rangle^{-1}\mathcal{P}(\langle\mathcal{C}\rangle^{-1})$ which we equate to $S_N=P_N(1)$ given by \eqref{star:DAN}. Using $\langle \mathcal{C}\rangle=AN^\mu$ gives
\begin{equation}
\label{Pz:small-A}
\mathcal{P}(z) \simeq 2 A^{1/\mu}\,z^{-1+1/\mu}\;,
\end{equation}
thus establishing \eqref{Pz:small}.

In contrast to the classic models of growing trees, the probability to create a star in the DAN, PAN, and IR networks decays very slowly with $N$. This observation provides a clue that typical network realizations grown by these rules should have non-negligible portions that are star-like, as seen in Fig.~\ref{fig:vis}.

\begin{figure}[ht]
    \centering
    \includegraphics[width=0.25\textwidth]{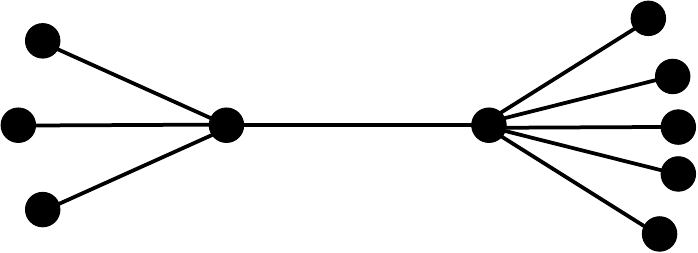}
    \caption{The two-hub (3,5) graph.}
    \label{fig:hmn}
\end{figure}

It is worth pointing out that the small-$z$ tail of $\mathcal{P}(z)$ depends on the initial condition.  To see how this dependence arises, consider the growth of a network starting with the linear graph $L_4$ with 4 nodes and core size $\mathcal{C}=2$ instead of the linear graph $L_3$, which was our canonical initial condition. Since the size of the core can only increase, the minimal core size remains equal to $2$ as the network grows. A graph with core size $2$ is necessarily a two-hub tree (Fig.~\ref{fig:hmn}). We denote by $H_{m,n}$ the probability to generate a two-hub $(m,n)$ graph, that is, a tree with $m$ leaves attached to one hub, $n$ leaves attached to the other hub, and a single link between the hubs. For the DAN, PAN, and VAN($\omega$)
models, the recurrence that determines the evolution of these probabilities when a new node attaches to the network is
\begin{eqnarray}
\label{H:LR}
H_{m,n} = \frac{m-1}{N-1}\,H_{m-1,n} + \frac{n-1}{N-1}\,H_{m,n-1}
\end{eqnarray}

Solving the first few of \eqref{H:LR} subject to the initial condition
$H_{1,1}=1$ corresponding to the linear graph $L_4$ gives
\begin{equation}
\begin{split}
& H_{2,1}=H_{1,2}=\frac{1}{4} \\
& H_{3,1}=H_{2,2}=H_{1,3}=\frac{1}{10} \\
& H_{4,1}=H_{3,2}=H_{2,3}=H_{1,4}=\frac{1}{20} ~.
\end{split}
\end{equation}
These probabilities clearly depend only on the total number of nodes:
$H_{m,n}=H_N$ for all $N=m+n+2$. This observation is readily proved by
induction. The recurrence \eqref{H:LR} then simplifies to $H_N=\frac{N-4}{N-1}\,H_{N-1}$ which we solve subject to the initial condition $H_4=1$ to give
\begin{align}
\label{HN}
H_N&=\frac{6}{(N-3)(N-2)(N-1)}\nonumber\\[2mm]
&=\frac{6}{(m+n-1)(m+n)(m+n+1)}\;,
\end{align}
for $N\geq 4$. The probability to build a two-hub tree of size $N$ is then
\begin{equation}
\label{two-hub}
P_N(2)=\sum_{m+n+2=N} H_{m,n} = \frac{6}{(N-2)(N-1)}\;.
\end{equation}

Specializing the core-size distribution in \eqref{Pz:scaled} to
$\mathcal{C}=2$, the minimal size of the core when the network starts with the graph $L_4$ is $P_N(2)=\langle \mathcal{C}\rangle^{-1}\mathcal{P}(2/\langle
\mathcal{C}\rangle)$. We now equate this to $P_N(2)$ in \eqref{two-hub} to give
\begin{equation}
\label{Pz:small-2}
\mathcal{P}(z) \simeq 12 \left(\frac{A}{2}\right)^{2/\mu}\,z^{-1+2/\mu}\;.
\end{equation}
The essential point is that $\mathcal{P}(z)$ vanishes as $z^{-1+2/\mu}$ when network growth starts with the graph $L_4$, rather than $\mathcal{P}(z)\simeq z^{-1+1/\mu}$ when growth begins with $L_3$---the small-$z$ tail of $\mathcal{P}(z)$ depends on the initial condition.

As a complement to the above presentation, we now focus on the large-$z$ tail of $\mathcal{P}(z)$.  We probe this limiting behavior by computing the probability $Q_N$ to create networks with the minimal number of leaves---namely, the linear graph of length $N$, $L_N$. We consider only the canonical initial condition where the network is $L_3$ so that $Q_3=1$. By straightforward enumeration, we obtain $Q_4=\frac{1}{3}$ and $Q_5=\frac{1}{6}$ for all of our redirection models.

For the PAN model, the probabilities $Q_N$ satisfy recurrence $Q_N=\frac{1}{2}Q_{N-1}$ from which
 \begin{equation}
\label{QN:0}
Q_N= \frac{2^{-(N-4)}}{3}\;.
\end{equation}
We now substitute
$Q_N=P_N(N\!-\!2)\simeq \langle \mathcal{C}\rangle^{-1} \mathcal{P}(N/\langle
\mathcal{C}\rangle)$ and $\langle \mathcal{C}\rangle=AN^\mu$ in the above equation and eliminate $N$ in favor of $z$.  This gives the large-$z$ asymptotic
\begin{equation}
\log \mathcal{P}(z) \simeq - \log 2\,(Az)^{1/(1-\mu)},
\end{equation}
up to the unknown amplitude $A$ in $\mathcal{C}\simeq A N^\mu$, as first quoted in Eq.~\eqref{Pz:large}.

For the VAN($\omega$) model with $\omega>0$, the recurrence satisfied by the probability $Q_N$ to build the line graph $L_N$ is $Q_{N+1}=\frac{2}{4+\omega(N-4)}Q_{N}$, from which
\begin{equation}
  \label{QN}
Q_N= \frac{1}{6}\left(\frac{2}{\omega}\right)^{N-5}\frac{\Gamma(1+\frac{4}{\omega})}{\Gamma(N-4+\frac{4}{\omega})}
\end{equation}
for $N\ge 5$. From this expression, we again obtain the same large-$z$ asymptotic behavior for $\mathcal{P}(z)$ as for DAN, namely, $\log\mathcal{P}(z)\sim -z^{\frac{1}{1-\mu}}\log z$, for any $\omega>0$.

\section{VAN$(\infty)$ model}
\label{sec:inf}

\begin{figure}
    \centering
\includegraphics[trim = -5 0 0 0, width=0.88\linewidth]{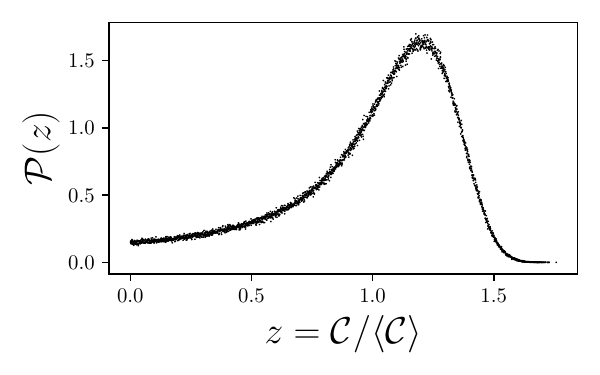}
\includegraphics[width=0.9\linewidth]{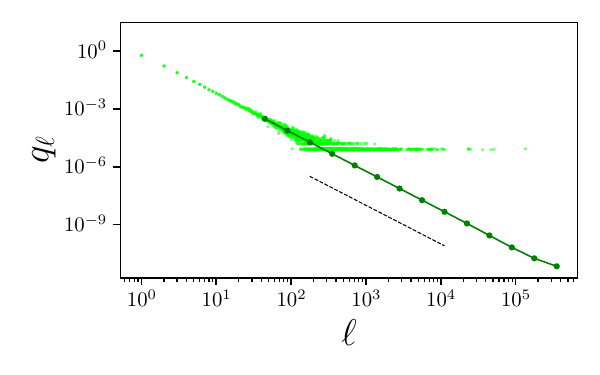}
    \caption{Scaled core size distribution $\mathcal{P}(z)$ and leaf degree distribution $q_\ell\sim \ell^{-2}$ in the VAN($\infty$) model. Data for the core size distribution are from $2.4\times 10^6$ realizations at size $N=10^4$, and for the leaf degree distribution are from $4800$ realizations at size $N=10^6$, with ratios $M_\ell/\mathcal{C}$ for three graphs shown as green dots; the dashed line represents pure powerlaw decay of the form $\ell^{-2}$.}
    \label{fig:VANinf_P_ldeg}
\end{figure}

The VAN($\infty$) model has unusual dynamics, residing at the boundary between linear and sublinear scaling. Since nodes in the nucleus have an infinite weight, whenever a nucleus node is created, it is immediately converted to a rank-$1$ node in the next attachment event. Thus the nucleus is either empty or consists of a single vertex. The limiting exponent for VAN($\infty$) is $\mu(\infty)=1$ with asymptotic approach \eqref{mu:omega-inf} as $\omega\rightarrow\infty$, from which we anticipate tail behavior $q_\ell\sim \ell^{-2}$. Equivalently, the leaf degree distribution is expected to scale as $M_\ell \sim \mathcal{C}/\ell^2$. We now recall the identity
\begin{align}
\label{N1-sum}
\sum_{\ell=1}^{N-1} \ell M_\ell=N_1\,.
\end{align}
The sum on the left-hand side of Eq.~\eqref{N1-sum} is dominated by the tail,
so that
\begin{equation}
\label{N1:LHS}
\sum_{\ell=1}^{N-1} \ell M_\ell \sim \mathcal{C}\sum_{\ell=1}^{N-1} \ell^{-1} \simeq \mathcal{C}\log N.
\end{equation}
Substituting this result, together with $N_1=N-\mathcal{C}$, into
Eq.~\eqref{N1-sum} we obtain
\begin{equation}
\label{C:inf}
\mathcal{C} \sim \frac{N}{\log N}.
\end{equation}
Thus the size of the core that is generated by VAN($\infty$) grows only logarithmically slower than $N$. Proceeding in the same heuristic manner, we deduce $q_\ell$ for all $\ell\geq 2$. Specializing \eqref{q-ell} to $\mu=1$ gives 
\begin{align}
\label{q-ell-inf}
(\ell+2)q_\ell  =  (\ell-1)q_{\ell-1}+q_{\ell+1}\,.
\end{align}
The minimal at infinity solution to this recurrence can be expressed via the generalized hypergeometric function  
\begin{equation}
\label{q:hyper-geom}
q_\ell  = A \ell^{-2} F[1,\ell-1; \ell+1, \ell+1;1]
\end{equation}
Higher-order hypergeometric functions can be expressed through integrals over the lower order ones. Specifically, we use the identity
\begin{equation*}
F[a,c; b, d;1]=\int_0^1 dt\,\frac{t^{c-1}(1-t)^{d-c-1}\Gamma(d)}{\Gamma(c)\Gamma(d-c)}\,F[a;b;t]
\end{equation*}
and recast Eq.~\eqref{q:hyper-geom} into
\begin{equation}
\label{q:int-hyper}
q_\ell  = A \frac{\ell-1}{\ell} \int_0^1 dt\, t^{\ell-2}(1-t)\,F[1;\ell+1;t]
\end{equation}
where $F[1;\ell+1;t]$ is the confluent hypergeometric function. Computing the integral we obtain
\begin{align}
\label{q-ell-inf:sol}
q_\ell  =  q_1[1-e(\ell-1)\,\gamma(\ell,1)]
\end{align}
where $\gamma(\ell,1)=\int_0^1 du\,u^{\ell-1}e^{-u}$ is the incomplete gamma function. Equation \eqref{q-ell-inf:sol} leads to the expected asymptotic behavior, $q_\ell\simeq q_1\ell^{-2}$. The amplitude $q_1$ is computed by normalization:
\begin{equation}
    \begin{aligned}
    \frac{1}{q_1}&=e\sum_{\ell=1}^{\infty}[\gamma(\ell,1)-\gamma(\ell+1,1)]=e\gamma(1,1)=e-1.
    \end{aligned}
\end{equation}
The solution $q_\ell$ can be written as an integral expression,
\begin{equation}
\label{q-ell-inf:sol2}
    q_\ell=\frac{1}{e-1}\int_0^1 dt \, e^{1-t}(1-t)t^{\ell-1}
\end{equation}
Exact expressions for $q_\ell$ for any $\ell\geq 1$ can then be extracted. For instance,
\begin{equation*}
q_2=\frac{3-e}{e-1}, \quad q_3=\frac{11-4e}{e-1}, \quad q_4=\frac{49-18e}{e-1},
\end{equation*}
suggesting the general form
\begin{subequations}
\label{q:AB}
\begin{equation}
q_\ell=\frac{A_\ell-B_\ell e}{e-1}
 \end{equation}
 with integer $A_\ell$ and $B_\ell$ for all $\ell\geq 2$. The general expressions for $A_\ell$ and $B_\ell$ are
\begin{equation}
A_\ell=\sum_{j=0}^{\ell-1} (j+1)!\binom{\ell-1}{j}\,, \quad B_\ell =\ell!-(\ell-1)!.
\end{equation}
\end{subequations}
The exact representation \eqref{q:AB} is compact, but, in contrast to
\eqref{q-ell-inf:sol} or \eqref{q-ell-inf:sol2}, not useful for extracting the asymptotic behavior. The scaled core-size distribution and leaf degree distribution for VAN($\infty$) are displayed in Fig.~\ref{fig:VANinf_P_ldeg}.

\section{Two-population model}
\label{sec:2p}

We have presented a self-contained theory for the phenomena of leaf proliferation, the ultra-broad leaf degree distribution, and the vanishingly small relative size of the core, in the family of random growing trees VAN$(\omega)$, emphasizing $\omega=0$, $\omega=1$, and $\omega=\infty$. To provide further intuition for these unusual properties, we analyze a toy model that sheds light on how leaf proliferation and a vanishingly small relative size of the core can arise.  Our discussion here serves as a natural counterpoint to that given in Sec.~\ref{sec:star} about the relative abundance of star graphs and reinforces the message that redirection models naturally lead to leaf proliferation and a vanishingly small core.

In our toy model, we partition the network into two populations, leaf nodes and core nodes, and disregard all other details, such as the degree of the core nodes. The number of leaf nodes and core nodes are $N_1$ and $\mathcal{C}$, respectively, with $N_1+\mathcal{C}=N$. Because these two populations are not independent, it suffices to consider only the dynamics of the core nodes. When a new node is added to the network only two outcomes can occur:
\begin{enumerate}
\itemsep -0.35 ex
\item A core node is added with probability $\mu \mathcal{C}(N)/N$.
\item Otherwise, a leaf node is added.
\end{enumerate}

These two steps are meant to mimic the isotropic redirection growth rule in which a core node is initially selected with probability $\mathcal{C}/N$ and then in a fraction $\mu$ of these selections attachment to a leaf occurs which increases $\mathcal{C}$ by $1$. In terms of the unknown parameter $\mu$, the number $\mathcal{C}(N)$ of core nodes changes according to the stochastic equation
\begin{equation}
\label{CN}
\mathcal{C}(N+1)=
\begin{cases}
\mathcal{C}(N)+1     & \text{prob.}~~\frac{\mu \mathcal{C}(N)}{N} \\
\mathcal{C}(N)        & \text{prob.}~~1- \frac{\mu \mathcal{C}(N)}{N}\;,
\end{cases}
\end{equation}
for $N\geq 3$. Starting with this exact equation we now compute moments of $\mathcal{C}$ and determine the probability distribution of $\mathcal{C}$.

\subsection{The average and the variance}

The number $\mathcal{C}(N)$ of core nodes is a random quantity for all $N\geq 4$. Averaging \eqref{CN} gives
\begin{equation}
\label{C:av}
\langle \mathcal{C}(N+1) \rangle=\left[1+\frac{\mu}{N}\right]\langle \mathcal{C}(N)\rangle\,.
\end{equation}
Iterating this recurrence and using the initial condition $\mathcal{C}(3)=1$, we obtain
\begin{equation}
\label{C:av-sol}
\langle \mathcal{C}(N)\rangle=A\,\frac{\Gamma(N+\mu)}{\Gamma(N)} \qquad A = \frac{2}{\Gamma(3+\mu)}\,.
\end{equation}
From the asymptotics of the gamma function, we have
\begin{equation}
\label{C:av-asymp}
\langle \mathcal{C}(N)\rangle=A\,N^\mu-\frac{\mu (1-\mu)A}{2}\,N^{\mu-1}+\ldots
\end{equation}
This toy two-population model thus predicts sublinear growth of $\mathcal{C}(N)$ on $N$, but also gives a simple mechanism, embodied by the process in Eq.~\eqref{CN}, for this anomalously slow growth.

In the same way as for the average, the recurrence for the second moment is
\begin{equation*}
\label{C2:av}
\langle \mathcal{C}^2(N+1) \rangle = \left[1+\frac{2\mu}{N}\right]\langle \mathcal{C}^2(N)\rangle
+ \mu A\,\frac{\Gamma(N+\mu)}{\Gamma(N+1)}\;,
\end{equation*}
from which
\begin{equation}
\label{C2:av-sol}
\langle \mathcal{C}^2(N)\rangle = \frac{4}{\Gamma(3+2\mu)}\,\frac{\Gamma(N+2\mu)}{\Gamma(N)}
- A\,\frac{\Gamma(N+\mu)}{\Gamma(N)}\;.
\end{equation}
The variance $\sigma^2(N)=\langle \mathcal{C}^2(N)\rangle - \langle
\mathcal{C}(N)\rangle^2$ thus has the asymptotic behavior
\begin{equation}
\label{sigma-asymp}
\sigma^2(N)=\left[\frac{4}{\Gamma(3+2\mu)}-A^2\right]N^{2\mu} - A N^\mu + \ldots
\end{equation}
The crucial implication is that the standard deviation is of the same order as the average. Thus different realizations of the growth process lead to differences in $\mathcal{C}(N)$ that are of the order of $\mathcal{C}(N)$ itself. That is, there is no asymptotic self-averaging in the number of core nodes, as we have observed in numerical simulations of networks grown by the DAN and PAN rules.

\subsection{The probability distribution $P_N(C)$}

The leading asymptotic behaviors of the average and variance,
Eqs.~\eqref{C:av-asymp} and \eqref{sigma-asymp}, imply that the random variable $\mathcal{C}(N)$ remains broadly distributed as $N\to\infty$. We therefore expect that the probability distribution $P_N(C):=\mathrm{Prob}[\mathcal{C}(N)=C]$ will converge to the scaling form
\begin{equation}
\label{P:scaling}
P_N(C) \simeq N^{-\mu}\,\mathcal{P}(z), \qquad z=C/N^\mu
\end{equation}
in the limit $C,N\to\infty$, with $z=C/N^\mu$ finite. The lack of the asymptotic self-averaging implies that the scaled distribution $\mathcal{P}(z)$ approaches a fixed finite-width shape.

Following the same reasoning as that used for the first two moments of
$C$, the probability distribution satisfies
\begin{align}
\label{PNC:rec}
P_{N+1}(C) = &\left[1- \frac{\mu C}{N}\right]P_N(C) \nonumber \\
&\vspace{3cm}+ \frac{\mu (C-1)}{N}\,P_N(C-1)
\end{align}
Let us first derive the asymptotic behaviors of $\mathcal{P}(z)$:
\begin{subequations}
\begin{align}
\label{P0}
&\mathcal{P}(0) =\frac{2}{\Gamma(3-\mu)} \\
\label{P-inf}
&\log \mathcal{P}(z) \sim - z^{1/(1-\mu)} \quad (z\to\infty)
\end{align}
\end{subequations}

To prove \eqref{P0}, we determine $P_N(1)$, i.e., the probability to
generate the star graph. Using \eqref{CN} we obtain
\begin{equation}
\label{PN1}
P_N(1)=\prod_{k=3}^{N-1}\left[1- \frac{\mu}{k}\right]=\frac{2}{\Gamma(N)}\,\frac{\Gamma(N-\mu)}{\Gamma(3-\mu)}
\end{equation}
For $N\to\infty$, $P_N(1)\simeq \frac{2}{\Gamma(3-\mu)}\,N^{-\mu}$,
which is consistent with \eqref{P:scaling} if $\mathcal{P}(0)$ is given by \eqref{P0}.

Let us now determine the probability to generate the opposite extreme of a line graph, namely $P_N(N-2)$. Again using \eqref{CN} we obtain
\begin{equation}
\label{PNN2}
P_N(N-2)=\prod_{k=3}^{N-1}\frac{\mu(k-2)}{k}=\frac{2\mu^{N-3}}{(N-1)(N-2)}
\end{equation}
Therefore, $\log P_N(N-2)\simeq \log \mathcal{P}(N^{1-\mu})\simeq N
\log\mu$. This leading behavior is consistent with \eqref{P-inf}, and
suggests that $\log \mathcal{P}(z) \simeq  z^{1/(1-\mu)}\,\log \mu$ as
$z\to\infty$.

For the extreme case of the maximum possible value $\mu=1$, rather special behavior arises. In this case, the extreme events of $\mathcal{C}=1$ and $\mathcal{C}=N-2$ occur with probabilities
\begin{equation*}
P_N(1)=\frac{2}{N-1} \quad \text{and} \quad P_N(N\!-\!2)=\frac{2}{(N-1)(N-2)}~.
\end{equation*}

A natural guess for $P_N(C)$ that interpolates between these two extreme values is 
\begin{equation}
\label{PNC:1}
P_N(C)=\frac{2}{N-1}\left[1-\frac{C-1}{N-2}\right]\;.
\end{equation}
One can verify that Eq.~\eqref{PNC:1} is indeed the solution by substituting this form into \eqref{PNC:rec} with $\mu=1$. Solvability of the stochastic process \eqref{CN} with $\mu=1$ is not surprising, as it reduces to the P\'{o}lya urn process \cite{urn-Mahmoud}.

The exact probability distribution \eqref{PNC:1} can be rewritten in the scaling form \eqref{P:scaling} with $\mu=1$, thereby justifying the emergence of scaling. The scaled distribution reads 
\begin{equation}
\label{Pz:1}
\mathcal{P}(z) = 
\begin{cases}
2(1-z)  & 0 < z <1\\
0          & z\geq 1
\end{cases}
\end{equation}

The scaled distribution depends on the initial condition. For our general models, we quantitatively exemplified this phenomenon by computing the behavior of $\mathcal{P}(z)$ in the $z\to 0$ limit for the default initial condition $L_3$ and the initial condition $L_4$. (The linear graph $L_4$ emerges from $L_3$ with probability $\frac{1}{3}$ for the VAN($\omega$) models independently of $\omega$.) The scaled distribution vanishes as $z^{\frac{1}{\mu}-1}$ for the initial condition $L_3$ and as $z^{\frac{2}{\mu}-1}$ for the initial condition $L_4$; cf. \eqref{Pz:small} and \eqref{Pz:small-2}. For the two-population model with $\mu=1$, the initial condition $L_4$ corresponds to $\mathcal{C}=2$ when $N=4$, and the scaled form of the solution of the P\'{o}lya urn process with this initial condition reads 
\begin{equation}
\label{Pz:2}
\mathcal{P}(z) = 
\begin{cases}
6z(1-z)  & 0 < z <1\\
0          & z\geq 1
\end{cases}
\end{equation}

We finally derive the exact solution in the $0<\mu<1$ range. We already know $P_N(1)$,  Eq.~\eqref{PN1}. Specializing \eqref{PNC:rec} to $C=2$ and using \eqref{PN1} we obtain an inhomogeneous recurrence 
\begin{equation}
\label{PN2:rec}
P_{N+1}(2) = \left[1- \frac{2\mu}{N}\right]P_N(2)+\frac{2\mu\Gamma(N-\mu)}{\Gamma(N+1)\Gamma(3-\mu)}
\end{equation}
which is solved (subject to $P_3(2)=0$) to yield
\begin{equation}
\label{PN2}
P_N(2)=\frac{2}{\Gamma(N)}\left[\frac{\Gamma(N-\mu)}{\Gamma(3-\mu)}-\frac{\Gamma(N-2\mu)}{\Gamma(3-2\mu)}\right].
\end{equation}
Specializing \eqref{PNC:rec} to $C=3$ and using \eqref{PN2} we obtain an inhomogeneous recurrence 
\begin{equation}
    \begin{aligned}
\label{PN3:rec}
P_{N+1}(3)& = \left[1- \frac{2\mu}{N}\right]P_N(3)  \\
&+ \frac{4\mu}{\Gamma(N+1)}\left[\frac{\Gamma(N-\mu)}{\Gamma(3-\mu)}-\frac{\Gamma(N-2\mu)}{\Gamma(3-2\mu)}\right]
\end{aligned}
\end{equation}
which is solved (subject to $P_4(3)=0$) to yield
\begin{equation*}
\label{PN3}
P_N(3)=\frac{2}{\Gamma(N)}\left[\frac{\Gamma(N-\mu)}{\Gamma(3-\mu)}-2\frac{\Gamma(N-2\mu)}{\Gamma(3-2\mu)}+\frac{\Gamma(N-3\mu)}{\Gamma(3-3\mu)}\right].
\end{equation*}

The exact results for $P_N(C)$ with $C=1,2,3$ suggest the general solution
\begin{equation}
\label{PNC}
P_N(C)=\frac{2}{\Gamma(N)}\sum_{k=1}^C (-1)^{k-1} \binom{C-1}{k-1}\frac{\Gamma(N-k\mu)}{\Gamma(3-k\mu)}
\end{equation}
applicable to $C\geq 1$ and all $N\geq C+2$. The solution \eqref{PNC} can be verified by direct substitution into \eqref{PNC:rec}. 

Unfortunately, it is difficult to extract useful information from the exact solution \eqref{PNC}. Since $\langle \mathcal{C}\rangle\sim N^\mu$, the core is large, so extracting the scaled distribution from \eqref{PNC} requires simplifying the sum of a very large number of terms. As another example we mention that extracting $P_N(N-2)$ from \eqref{PNC} requires summing $N-2$ terms, while directly from the recurrence \eqref{PNC:rec} we deduced \eqref{PNN2}.

\bibliography{references-nets.bib}

\end{document}